\documentclass
[superscriptaddress,secnumarabic,amssymb,amsmath,nobibnotes,aps,prd,showkeys,showpacs,nofootinbib,onecolumn,notitlepage,12pt]{revtex4}%
\usepackage{setspace}
\usepackage{color}
\usepackage{amsmath}
\usepackage{amsfonts}
\usepackage{verbatim}
\usepackage{amssymb}
\usepackage{graphicx,bm}
\usepackage{amsmath}
\usepackage{amssymb}
\usepackage{graphicx}
\usepackage[caption=false]{subfig}%
\providecommand{\U}[1]{\protect\rule{.1in}{.1in}}
\newcommand{\be}{\begin{equation}}
\newcommand{\ee}{\end{equation}}

\newcommand{\mincir}{\raise
-3.truept\hbox{\rlap{\hbox{$\sim$}}\raise4.truept\hbox{$<$}\ }}
\newcommand{\magcir}{\raise
-3.truept\hbox{\rlap{\hbox{$\sim$}}\raise4.truept\hbox{$>$}\ }}

\ifx\pdfoutput\relax\let\pdfoutput=\undefined\fi
\newcount\msipdfoutput
\ifx\pdfoutput\undefined\else
\ifcase\pdfoutput\else
\ifx\paperwidth\undefined\else
\ifdim\paperheight=0pt\relax\else\pdfpageheight\paperheight\fi
\ifdim\paperwidth=0pt\relax\else\pdfpagewidth\paperwidth\fi
\fi\fi\fi
\begin{document}
\title{$f\left(  R,\square R\right)  $-gravity and equivalency with the modified GUP
Scalar field models}
\author{Andronikos Paliathanasis}
\email{anpaliat@phys.uoa.gr}
\affiliation{Institute of Systems Science, Durban University of Technology, Durban 4000,
South Africa}
\affiliation{Departamento de Matem\'{a}ticas, Universidad Cat\'{o}lica del Norte, Avda.
Angamos 0610, Casilla 1280 Antofagasta, Chile}

\begin{abstract}
Inspired by the generalization of scalar field gravitational models with a
minimum length we study the equivalent theory in modified theories of gravity.
The quadratic Generalized Uncertainty Principle (GUP) gives rise to a deformed
Heisenberg algebra in the application, resulting in the emergence of
additional degrees of freedom described by higher-order derivatives. The new
degrees of freedom can be attributed to the introduction of a new scalar
field, trasforming the resulting theory into a two-scalar field theory. Thus,
in order to describe all the degrees of freedom we investigate special forms
of the sixth-order modify $f\left(  R,\square R\right)  -$theory of gravity,
where the gravitational Lagrangian has similar properties~to~that of the GUP
scalar field theory. Finally, the cosmological applications are discussed, and
we show that the de Sitter universe can be recovered without introducing a
cosmological constant.

\end{abstract}
\maketitle
\date{\today}

\section{Introduction}

\label{sec1}

Dark energy is an exotic matter source introduced in Einstein's field
equations of General Relativity in order to explain the cosmic acceleration as
it is observed by the cosmological data \cite{rr1,Teg,Kowal,Komatsu,suzuki11}.
Currently, there are no observable phenomena directly linked to the nature and
characteristics of dark energy. As a result, the physical nature and the
origin of dark energy it is up for debate in the scientific community.
Introducing the cosmological constant in the Einstein-Hilbert Action Integral
represents one of the simplest approaches to address the issue of dark energy.
The $\Lambda$CDM cosmology is an analytic solution of the field equations of
General Relativity for a spatially flat
Friedmann--Lema\^{\i}tre--Robertson--Walker (FLRW) geometry with a nonzero
cosmological constant term and a pressureless dust fluid. $\Lambda$CDM has
achieved considerable success in describing a large span of astronomical and
cosmological data. Nevertheless, the cosmological constant can not explain the
complete evolution of the cosmic history \cite{l01,l02}, we refer the reader
to the recent discussion \cite{l03}.

In order to overcome the problems of the cosmological constant and to explain
the late-time cosmic acceleration, in recent years, cosmologists have proposed
various models which can categorized to two large families of theories, the
dark energy theories and the modified theories of gravity. For the dark energy
models, an energy-momentum tensor that attributes the new degrees of freedom
is introduced in field equations of General Relativity. The dynamics driven by
the newly introduced degrees of freedom provide an explanation for various
cosmological phenomena. Some of the most common dark energy models are the
quintessence scalar field \cite{ratra,q5,qq1,qq4}, phantom scalar field
\cite{q22,q23}, chameleon mechanism \cite{qq01}, scalar-tensor models
\cite{st1,st2,st3}, Galileons \cite{st4,st5}; multi-scalar field models
\cite{st6,st7,st8}, Chaplygin gas-like fluids \cite{q20,q21,q21a,q21b,q21c},
k-essence \cite{q26,q26a}, tachyons \cite{q25,q25a,q25b}. On the other hand,
in modified theories of gravity the Einstein-Hilbert Action Integral is
modified with the introduction of geometric invariants \cite{md1,md2,md3}. As
a result, the gravitational field equations are modified such that new
geometrodynamical components to be introduced and provide an effective
geometric matter source to explain the acceleration of the universe
\cite{cp1}. The "zoology" of modified theories of gravity can be categorized
based on the geometric invariant used to modify the gravitational Action
Integral and the order of derivatives involved. Within the realm of modified
theories of gravity, a particular family of interest comprises the so-called
$f\left(  X\right)  ~$-theories, where the gravitational Action Integral is a
function $f$ of the geometric invariant $X$. The latter can be the Ricci
scalar leads to $f\left(  R\right)  ~$-gravity \cite{Buda}, the torsion scalar
$T$ of the Weitzenb\"{o}ck connection in teleparallelism \cite{Ferraro}, the
non-metricity scalar $Q$ in symmetric teleparallel theory \cite{f6}, the
Gauss-Bonnet term \cite{bb1}, for other proposed theories see for instance
\cite{bb2,bb3,bb4,bb5,bb6,bb7} and references therein.

$f\left(  R\right)  ~$-gravity \cite{Sotiriou,odin1} has been widely studied
in the literature with many interesting results in cosmological studies
\cite{cf1,cf2,cf3} as also important results in the description of
astrophysical objects \cite{cf4,cf5,cf6}. The quadratic theory of gravity
$f\left(  R\right)  ~=R+qR^{2}$ \cite{qua1,qua2,qua3}, has been used
successfully \cite{planck2015} for the description of another inflationary
epoch of our universe. Specifically, the quadratic $f\left(  R\right)
$-gravity has been used as a mechanism for acceleration in the early stages of
the universe \cite{star}. The quadratic term $R^{2}$ follows from the analytic
expressions for the quantum-gravitational effects in the one-loop
approximation. Indeed the origin of the $R^{2}$ term is the vacuum
polarization$~$of the physical space \cite{df1,df2}.

The existence of a minimum length, i.e. maximum energy in nature, is supported
by various approaches to quantum gravity, like string theory, doubly special
relativity and the black hole physics. The incorporation of a minimum length
necessitates a modification of Heisenberg's Uncertainty Principle, giving rise
to the Generalized Uncertainty Principle (GUP) \cite{Maggiore}. By modifying
the Uncertainty Principle governing quantum observables, we arrive at a
revised definition of the Heisenberg algebra, which in turn leads to
adjustments in the Poisson brackets in the classical limits.
\cite{cas1,cas2,cas3}.\ The modified Poisson brackets revise the equations of
motion such that new degrees of freedom to be introduced. There is a plethora
of studies of GUP in gravitational physics, see for instance
\cite{mc1,mc2,mc3}. As far as the cosmological constant is concerned, it has
been found that it is related to the GUP and the minimum length \cite{cc1,cc2}%
. Another extension of the Uncertainty Principle in the Extended Uncertainty
Principle (EUP) \cite{ep1,ep2}, however in this work we are focus in GUP. We
refer the reader to the recent review \cite{gup11}

The quintessence model, employed to describe dark energy in \cite{angup1}
incorporates modifications to the scalar field Lagrangian through the
application of the quadratic GUP, that is,\ in the equation of motion for the
scalar field, i.e. the Klein-Gordon equation, new higher-order derivatives
have been introduced, as a result of the deformed Heisenberg algebra. These
new terms are related to the existence of the minimum length. The higher-order
derivatives can be described by a second-scalar field and this modified
quintessence model is equivalent to a multi-scalar field cosmological model
with interaction between the two scalar fields. It was found that the GUP
components affect the cosmological evolution of a FLRW\ geometry not only in
the early stages of the universe but also in the late-universe, while the
effects for the existence of the deformed Heisenberg algebra are observable in
the cosmological perturbations \cite{angup2}. The case where a matter source
is included in the field equations coupled to the scalar field was the subject
of study in \cite{angup3}. Furthermore, in \cite{angup4} the deformed
Heisenberg algebra for the quadratic GUP was considered to study the effects
of the minimum length in the case of a gravitational theory, which satisfies
Mach's Principle. Specifically, the case of the Bran-Dicke scalar field was
considered modified by the quadratic GUP. It was found that the modified
theory is a multi-scalar field theory with different dynamics and evolution
from the unmodified theory. An important result is that the nature of the
asymptotic solutions~does~not depend on the value of the Brans-Dicke
parameter, which is different from the case of the unmodified theory. Last but
not least, the effects of the GUP modification in the Brans-Dicke theory are
observable not only in the early stages of the universe but also in the late-time.

There exists a unique connection between some modified theories of gravity and
scalar field models. Indeed, the new degrees of freedom provided by the
geometric scalars in modified theories of gravity can be attributed to scalar
fields. For instance $f\left(  R\right)  ~$-gravity \cite{Sotiriou} is
equivalent with the so-called O'Hanlon theory \cite{ohan} which belongs to the
family of scalar-tensor theories \cite{far1} and specifically to the
Brans-Dicke gravity \cite{brans}. Recall that $f\left(  R\right)  $-gravity is
fourth-order theory for a nonlinear function $f$, while it reduces to a
second-order theory in the limit of General Relativity when $f$ is a linear
function. Thus, with the introduction of a Lagrange multiplier the field
equations can be written in the equivalent order of second-order derivatives
but in the same time the number of the dependent variables increases. In
general, Lagrange multipliers are applied for the introduction of constraints
for the cosmological model. There are various studies in the literature for
the application of Lagrange multipliers in gravitational theories, see for
instance \cite{lan1,lan2,lan3}.

In this study, we investigate if there exists a modified theory of gravity
where we can recover the quadratic GUP corrections that follow from the
deformation of the Heisenberg algebra in scalar field theories. With such
analysis, we will be able to find the geometric equivalent of GUP in modified
theories of gravity. The GUP scalar field models studied before, in the
Einstein and Jordan frames \cite{angup1,angup4}, are second-order multi-scalar
field models, specifically the matter source is attributed to two-scalar
fields. Hence, in order to be able to recover a two scalar field theory in
modified theories of gravity we shall consider a sixth-order theory. $F\left(
R,\square R\right)  ~$-theory has been introduced before \cite{sx1,sx2} as a
sixth-order gravity and extension of $f\left(  R\right)  ~$ theory, where
$\square$ is the Laplace operator. The introduction of higher-order
derivatives in the gravitational Action Integral is in agreement with quantum
gravity \cite{sx1}. The effects of the $\square R$ terms in $F\left(
R,\square R\right)  ~-$theory has been widely studied before in the
description of inflation \cite{sx3,sx4,sx4a}. A detailed analysis of the
cosmological dynamics in $F\left(  R,\square R\right)  $ performed recently in
\cite{sx5} where it was found that higher-order terms can dominate the
evolution of the universe. For more applications of $F\left(  R,\square
R\right)  ~$-theory in gravitational physics we refer the reader to
\cite{sx6,sx7,sx8,sx9,sx10} and references therein. The structure of the paper
is as follows.

In Section \ref{GUPmd} we present the basic properties and definitions of GUP
and we focus on the case of quadratic GUP. We define the deformed Heisenberg
algebra, and we derive the modified Klein-Gordon equation for a spin-0
particle. The latter modified Lagrangian is used in Section \ref{sec3} in
order to introduce the effects of the deformed Heisenberg algebra in scalar
field cosmological models. The modified quintessence and modified Brans-Dicke
models are presented. In Section \ref{sec4} we consider the $F\left(
R,\square R\right)  $-gravity which is a theory of gravity of sixth-order. We
introduce Lagrange multipliers in order to increase the number of the
dependent variables with the introduction of scalar fields, and at the same
time reduce the theory into a second-order gravitational model. We found that
a separable function $F\left(  R,\square R\right)  ~$ is equivalent to
two-scalar field theory with similar properties to that of GUP scalar field
models. In Section \ref{sec5} we focus in the case of $F\left(  R,\square
R\right)  ~=R+K\left(  ~\square R\right)  ~$ gravity, where the term $K\left(
~\square R\right)  ~$ introduces similar corrections terms in the field
equations as that of the minimum length of GUP. Indeed, the $K\left(  ~\square
R\right)  ~$ we can say that follows from the deformation algebra of the
Einstein-Hilbert Action Integral. For a spatially flat FLRW background
geometry~in Section \ref{sec5} it was found that the de Sitter universe is a
unique attractor for the cosmological solution without necessary introduce a
cosmological constant term. Thus, the existence of a minim length in the early
universe leads to an accelerated universe. Finally, in Section \ref{sec6} we
draw our conclusions.

\section{Generalized Uncertainty Principle}

\label{GUPmd}

The existence of a minimum length leads to the modification of the
Heisenberg's uncertainty principle as
\begin{equation}
\Delta X_{i}\Delta P_{j}\geqslant\frac{\hbar}{2}[\delta_{ij}(1+\beta
P^{2})+2\beta P_{i}P_{j}], \label{GUP}%
\end{equation}
where parameter $\beta$ is the deformed parameter defined as $\beta={\beta
_{0}}/{M_{Pl}^{2}c^{2},}$ where $M_{Pl}$ is the Planck mass and $M_{Pl}c^{2}$
is the Planck energy, or equivalently $\beta={\beta_{0}\ell_{Pl}^{2}}%
/{\hbar^{2}}$ where $\ell_{Pl}$ ($\approx10^{-35}~m)$ is the Planck length.
Parameter $\beta_{0}$ usually is selected to be positive and equal to one,
however in order to have observable quantum effects, the parameter $\beta_{0}$
can have different values \cite{Vagenas}.

\bigskip

The usual choice of the parameter $\beta_{0}$ is $\beta_{0}=1$, however, such
a choice could lead do not observable quantum effects; however in
\cite{Vagenas}, it has been shown that the dimensionless parameter $\beta_{0}$
could has upper bound such as $\beta_{0}>>1$. However, there are studies where
shown that the deformation parameter $\beta_{0}$ can be negative
\cite{neg1,neg2,neg5,neg6}.

Therefore, the modified uncertainty principle (\ref{GUP}) leads to the
deformed Heisenberg algebra \cite{Kemph1,Kemph2}
\begin{equation}
\lbrack X_{i},P_{j}]=i\hbar\lbrack\delta_{\alpha\beta}(1+\beta_{0}\frac
{\ell_{Pl}^{2}}{2\hbar^{2}}P^{2})+\beta_{0}\frac{\ell_{Pl}^{2}}{\hbar^{2}%
}P_{\alpha}P_{\beta}]. \label{xp-com}%
\end{equation}
Hence, the coordinate representation of the momentum operator, which satisfies
the commutation relation (\ref{xp-com}), can be defined as $P_{i}%
=p_{i}(1+\beta p^{2}),$ where we have selected to keep underformed the
position underformed, that is, $X_{i}=x_{i}$. Representation $(x,p)$ is the
canonical representation satisfying $[x_{i},p_{j}]=i\hbar\delta_{ij}$.\newline

In the relativistic limit the commutation relation (\ref{xp-com}) reads
\cite{Moayedi}
\begin{equation}
\lbrack X_{\mu},P_{\nu}]=-i\hbar\lbrack(1+\beta_{0}\frac{\ell_{Pl}^{2}}%
{2\hbar^{2}}(\eta^{\mu\nu}P_{\mu}P_{\nu}))\eta_{\mu\nu}+\beta_{0}\frac
{\ell_{Pl}^{2}}{\hbar^{2}}P_{\mu}P_{\nu}], \label{eqn3}%
\end{equation}
where $\eta_{\mu\nu}$ is the flat metric. Therefore, the corresponding
deformed operators for (\ref{eqn3}) are $P_{\mu}=p_{\mu}(1-\beta(\eta
^{\alpha\gamma}p_{\alpha}p_{\gamma}))~,~~X_{\nu}=x_{\nu}.$

In the relativistic limit the equation of motion for a spin-0 particle with
rest mass zero is
\begin{equation}
\left[  \eta^{\mu\nu}P_{\mu}P_{\nu}-\left(  mc\right)  ^{2}\right]  \Psi=0,
\end{equation}
that is,
\begin{equation}
\square\Psi-2\beta\hbar^{2}\square\left(  \square\Psi\right)  +\left(
\frac{mc}{\hbar}\right)  ^{2}\Psi+O\left(  \beta^{2}\right)  =0. \label{eq33}%
\end{equation}
where the Laplace operator for the metric $\eta_{\mu\nu}$ is marked with the
symbol $\square$ . For a generic metric tensor $g_{\mu\nu}$ the Laplace
operator is defined as $\Delta=\square$, where $\Delta=\frac{1}{\sqrt{-g}%
}\partial_{\mu}\left(  g^{\mu\nu}\sqrt{-g}\partial_{\mu}\right)  $. In
equation (\ref{eq33}) term $2\beta\hbar^{2}\square\left(  \square\Psi\right)
$ is the quantum correction term which follows from the existence of the
minimum length. Because the quantum correction terms introduce the
fourth-order derivatives in the Klein-Gordon equation, the $\beta
^{2}\rightarrow0$, equation (\ref{eq33}) is a singular perturbative system
which means that there is a solution where the term $2\beta\hbar^{2}%
\square\left(  \square\Psi\right)  $ dominates and drives the dynamics. For
the discussion of inner and outer solutions of singular pertubative
differential equations we refer the reader in \cite{Tikhonov}.

The modified Klein-Gordon equation (\ref{eq33}) is a fourth-order partial
differential equation. It can be written in the equivalent form of two
second-order differential equations by introducing the new scalar field
$\Phi=\square\Psi$. Hence, equation (\ref{eq33}) in the limit $\beta
^{2}\rightarrow0$ becomes,%
\begin{equation}
\square\Psi-\Phi=0, \label{eq.01}%
\end{equation}%
\begin{equation}
2\beta\hbar^{2}\square\Phi+\left(  mc\right)  ^{2}\Psi+\Phi=0. \label{eq.02}%
\end{equation}

Let us now derive the Lagrangian for the two scalar field model with equations
of motion (\ref{eq.01}) and (\ref{eq.02}).

We consider the Action Integral of the equation (\ref{eq33})
\begin{equation}
S_{KG}^{\operatorname{mod}}=\int dx^{4}\left(  \frac{1}{2}\eta^{\mu\nu
}\mathcal{D}_{\mu}\Psi\mathcal{D}_{\nu}\Psi-\frac{1}{2}\left(  \frac{mc}%
{\hbar}\right)  ^{2}\Psi^{2}\right)  , \label{eq.03}%
\end{equation}
where now $\mathcal{D}_{\mu}$ is the deformed operator defined as%
\begin{equation}
\mathcal{D}_{\mu}=\nabla_{\mu}+\beta\hbar^{2}\nabla_{\mu}\square.
\end{equation}

Hence, expression (\ref{eq.03}) becomes%
\begin{equation}
S_{KG}^{\operatorname{mod}}=\int dx^{4}\left(  \frac{1}{2}\eta^{\mu\nu}\left(
\nabla_{\mu}\Psi\nabla_{\nu}\Psi+2\beta\hbar^{2}\left(  \nabla_{\nu}%
\nabla_{\mu}\square\right)  \Psi\right)  -\frac{1}{2}\left(  \frac{mc}{\hbar
}\Psi\right)  ^{2}\right)  ,
\end{equation}
We include the Lagrange multiplier $\lambda$ into the dynamical system and the
second scalar field $\Phi$ with constraint $\Phi=\square\Psi$, therefore the
latter Action Integral after integration by parts is written as follows%

\begin{equation}
S_{KG}^{\operatorname{mod}}=\int dx^{4}\sqrt{-g}\left(  \frac{1}{2}\eta
^{\mu\nu}\nabla_{\mu}\Psi\nabla_{\nu}\Psi+2\beta\hbar^{2}\eta^{\mu\nu}%
\nabla_{\mu}\Psi\nabla_{\nu}\Phi+\beta\hbar^{2}\Phi^{2}-\frac{1}{2}\left(
\frac{mc}{\hbar}\right)  ^{2}\Psi^{2}\right)  . \label{bd.06}%
\end{equation}

\section{Scalar field theories modified by GUP}

\label{sec3}

We review previous results on the application of the quadratic GUP in scalar
field theories.

Consider the Scalar-tensor Action Integral \cite{far1}%
\begin{equation}
S=\int dx^{4}\sqrt{-g}\left[  \frac{1}{2}F\left(  \phi\right)  ^{2}R-L_{\phi
}\right]  , \label{bd.04}%
\end{equation}
where $F\left(  \phi\right)  $ is the coupling function and $L_{\phi}$ is the
Lagrangian for the scalar field, that is
\begin{equation}
L_{\phi}\left(  x^{\mu},\phi,\nabla_{\mu}\phi\right)  =\frac{\omega}{2}%
g^{\mu\nu}\nabla_{\mu}\phi\nabla_{\nu}\phi+V\left(  \phi\right)  \label{bd.05}%
\end{equation}
where $\omega$ is a parameter which defines the nature of the scalar field.
Lagrangian function (\ref{bd.05}) is that of the underformed Heisenberg
algebra. Hence, after the application of the quadratic GUP, we end with
Lagrangian function (\ref{bd.06}) which reads%
\begin{equation}
L_{\phi}^{GUP}\left(  x^{\mu},\phi,\psi,\nabla_{\mu}\phi,\nabla_{\mu}%
\psi\right)  =\frac{\omega}{2}g^{\mu\nu}\nabla_{\mu}\phi\nabla_{\nu}%
\phi+V\left(  \phi\right)  +\beta\left(  2g^{\mu\nu}\nabla_{\mu}\phi
\nabla_{\nu}\psi+\psi^{2}\right)  \label{bd.07}%
\end{equation}
where $\psi$ is the second scalar field which follows from the higher-order derivatives.

We replace in (\ref{bd.07}) , (\ref{bd.04}) and we end with the Action
Integral%
\begin{equation}
S=\int dx^{4}\sqrt{-g}\left[  \frac{1}{2}F\left(  \phi\right)  R-\frac{\omega
}{2}g^{\mu\nu}\nabla_{\mu}\phi\nabla_{\nu}\phi-V\left(  \phi\right)
-\beta\left(  2g^{\mu\nu}\nabla_{\mu}\phi\nabla_{\nu}\psi+\psi^{2}\right)
\right]  . \label{bd.08}%
\end{equation}

\subsection{Quintessence and phantom cosmologies}

In the case of a spatially flat FLRW geometry
\begin{equation}
ds^{2}=-dt^{2}+a^{2}\left(  t\right)  \left(  dx^{2}+dy^{2}+dz^{2}\right)  ,
\label{flrw}%
\end{equation}
and for a minimally coupled scalar field, that is $F\left(  \phi\right)  =1$,
the cosmological field equations are
\begin{equation}
3H^{2}-\frac{1}{2}\dot{\phi}^{2}-2\beta\dot{\phi}\dot{\psi}-V\left(
\phi\right)  +\beta\psi^{2}=0, \label{ac.10}%
\end{equation}%
\begin{equation}
2\dot{H}+3H^{2}+\frac{1}{2}\dot{\phi}^{2}+2\beta\dot{\phi}\dot{\psi}-\left(
V\left(  \phi\right)  -\beta\psi^{2}\right)  =0, \label{ac.11}%
\end{equation}%
\begin{align}
\ddot{\phi}+3H\dot{\phi}-\psi &  =0,\label{ac.12a}\\
\beta\left(  \ddot{\psi}+3H\dot{\psi}\right)  +\frac{1}{2}\left(
\psi+V_{,\phi}\right)   &  =0. \label{ac.12b}%
\end{align}
where $H=\frac{\dot{a}}{a}$ is the Hubble function.

For $\omega=1$, the scalar field is a quintessence while for $\omega=-1$
scalar field $\psi$ is a phantom field. However, as it was found before in
\cite{angup1}, the equation of state parameter for the effective parameter can
cross the phantom divide line and for the case of a quintessence field, that
is, because the second scalar field $\psi$ can dominates such that
$w_{eff}<-1$. Last but not least, from the analysis of the dynamics in
\cite{angup2} it was found that the de Sitter universe is always a late-time
attractor independent from the nature of the scalar field potential.

Furthermore, the field equations (\ref{ac.10})-(\ref{ac.12b}) can derive from
the variation of the point-like Lagrangian
\begin{equation}
L_{Q}=3a\dot{a}^{2}-\frac{1}{2}a^{3}\dot{\phi}^{2}-2\beta a^{3}\dot{\phi}%
\dot{\psi}+a^{3}V\left(  \phi\right)  -\beta a^{3}\psi^{2}.
\end{equation}

\subsection{Brans-Dicke cosmology}

Brans-Dicke model is recovered when $F\left(  \phi\right)  =\phi^{2}$, where
$\omega=\bar{\omega}_{BD}$ is now the Brans-Dicke parameter. In this case the
cosmological field equations in the case of a FLRW spacetime%
\begin{equation}
6H^{2}+12H\left(  \frac{\dot{\phi}}{\phi}\right)  +\frac{\bar{\omega}_{BD}}%
{2}\left(  \dot{\phi}^{2}+2\beta\hbar^{2}\left(  \frac{\dot{\phi}}{\phi
}\right)  \left(  \frac{\dot{\psi}}{\phi}\right)  -\frac{\psi^{2}}{\phi^{2}%
}\right)  +\frac{V\left(  \phi\right)  }{\phi^{2}}=0,
\end{equation}%
\begin{equation}
2\dot{H}+3H^{2}+4\left(  \frac{\dot{\phi}}{\phi}\right)  H+\left(  \frac
{\bar{\omega}_{BD}}{4}-2\right)  \left(  \frac{\dot{\phi}}{\phi}\right)
^{2}+\frac{V\left(  \phi\right)  }{2\phi^{2}}+\beta\frac{\bar{\omega}_{BD}%
}{4\phi^{2}}\left(  \psi^{2}-2\dot{\psi}\dot{\phi}\right)  =0,
\end{equation}%
\begin{equation}
\bar{\omega}_{BD}\left(  \beta\ddot{\psi}+\ddot{\phi}\right)  +3\bar{\omega
}_{BD}H\left(  \beta\dot{\psi}+\dot{\phi}\right)  +V_{,\phi}+12\phi\left(
\dot{H}+2H^{2}\right)  =0.\label{pd.04}%
\end{equation}%
\begin{equation}
\ddot{\phi}+3H\dot{\phi}+\psi=0,
\end{equation}
Similarly as before the Lagrangian function which reproduces the field
equations reads%
\begin{equation}
L\left(  a,\dot{a},\phi,\dot{\phi},\psi,\dot{\psi}\right)  =-6a\phi^{2}\dot
{a}^{2}-12\phi a^{2}\dot{a}\dot{\phi}-\frac{\bar{\omega}_{BD}}{2}a^{3}\left(
\dot{\phi}^{2}+2\beta\hbar^{2}\dot{\phi}\dot{\psi}-\psi^{2}\right)
+a^{3}V\left(  \phi\right)  .\label{pd.66}%
\end{equation}

In \cite{angup4} it was found that the modified field equations provide a
different cosmological history. The GUP introduces significant changes to the
dynamics and asymptotic solutions of the field equations, resulting in
newfound degrees of freedom. These modifications diverge from those of the
unmodified model. Moreover, the physical characteristics of the asymptotic
solutions are contingent upon the exponent of the potential function rather
than the value of the Brans-Dicke parameter, which is the case in the
unmodified model.

\section{$F\left(  R,\square R\right)  $-gravity}

\label{sec4}

In this Section we incoporate Lagrange multipliers in order to study the
scalar-tensor description of $F\left(  R,\square R\right)  $-gravity. In our
study, we focus into the scenario where scalar field models with GUP are
derived from $F\left(  R,\square R\right)  $-gravity.

\subsection{$f\left(  R\right)  $-gravity}

The Ricciscalar $R$ is defined by second-order derivatives of the metric
tensor; hence a Lagrangian which is an arbitrary function of $R$ leads to
fourth-order equations of motions. Indeed, $f\left(  R\right)  $-gravity is a
fourth-order theory and equivalent with a special case of the scalar-tensor theory.

Consider a Riemannian manifold with metric $g_{\mu\nu}$ and Ricciscalar $R$.
The Action Integral of the $f\left(  R\right)  $-gravity is as follows%
\begin{equation}
S=\int dx^{4}\sqrt{-g}f\left(  R\right)  \label{gup.3}%
\end{equation}
when $f\left(  R\right)  =R-2\Lambda$ the limit of General Relativity is recovered.

We make use of the Lagrange multiplier $\lambda$, thus the Action Integral
(\ref{gup.3}) reads%
\begin{equation}
S=\int dx^{4}\sqrt{-g}\left(  f\left(  \chi\right)  +\lambda\left(
\chi-R\right)  \right)  . \label{gup.4}%
\end{equation}
The equation of motion for $\lambda$ is $\frac{\partial S}{\partial\chi}=0$,
that is, $\lambda=-f_{,\chi}$.

By replacing $\lambda$ in (\ref{gup.4}) we find%
\begin{equation}
S=\int dx^{4}\left[  f_{,\chi}R+\left(  f-\chi f_{,\chi}\right)  \right]  ,
\label{gup.5}%
\end{equation}
or equivalently%
\begin{equation}
S=\int dx^{4}\left[  \varphi R+V\left(  \varphi\right)  \right]
,~\varphi=f_{,\chi},
\end{equation}
and potential function $V\left(  \varphi\right)  =\left(  f-\chi f_{,\chi
}\right)  \,$.

\subsection{$F\left(  R,\square R\right)  $-gravity}

We consider the sixth-order theory of gravity with Action Integral%
\begin{equation}
S=\int dx^{4}\sqrt{-g}F\left(  R,\square R\right)  \label{gup.6}%
\end{equation}
where $\square R=g^{\mu\nu}R_{;\mu\nu}$.

We consider the new variables $\chi,\zeta$ and we introduce the Lagrange
multipliers $\lambda_{1},~\lambda_{2}$. Therefore, the Action Integral
(\ref{gup.6}) is expressed as follows%
\begin{equation}
S=\int dx^{4}\sqrt{-g}\left[  F\left(  \chi,\zeta\right)  +\lambda_{1}\left(
\chi-R\right)  +\lambda_{2}\left(  \zeta-\square R\right)  \right]  .
\label{gup.7}%
\end{equation}

Variation with respect to the Lagrange multipliers provide the equations of
motions $\lambda_{1}+F_{,\chi}=0$ and $\lambda_{2}+F_{,\zeta}=0$.

We replace in the gravitational Action Integral (\ref{gup.7}) and we find%
\begin{equation}
S=\int dx^{4}\sqrt{-g}\left[  \left(  F_{,\chi}+F_{,\zeta}\square\right)
R+\left(  F-\chi F_{,\chi}-\zeta F_{,\zeta}\right)  \right]  \label{gup.9}%
\end{equation}

Integration by parts of the second term of (\ref{gup.9}) gives
\[
\int dx^{4}\sqrt{-g}\left(  F_{,\zeta}\square R\right)  =-\int dx^{4}\sqrt
{-g}\left(  F_{,\zeta\zeta}g^{\mu\nu}\chi_{;\mu}\zeta_{;\nu}\right)  .
\]
Consequently we end with the Gravitational Action Integral
\begin{equation}
S=\int dx^{4}\sqrt{-g}\left[  F_{,\chi}R-F_{,\zeta\zeta}g^{\mu\nu}\nabla_{\mu
}\chi\nabla_{\nu}\zeta+\left(  F-\chi F_{,\chi}-\zeta F_{,\zeta}\right)
\right]  \label{gup.10}%
\end{equation}

As we expected we have two scalar fields, the $\left\{  \chi,\zeta\right\}  $,
because $F\left(  R,\square R\right)  $-gravity is a sixth-order gravity. The
field $\chi$ is a \ non-minimally coupled field and the field $\zeta$ is an
extra field which can be seen as perturbation effects. What is of special
interest is the non-diagonal term $F_{,\zeta\zeta}g^{\mu\nu}\chi_{;\mu}%
\zeta_{;\nu}$ which has similarities with the non-diagonal term in
(\ref{bd.08}).

\subsubsection{$F\left(  R,\square R\right)  =f\left(  R\right)  +K\left(
\square R\right)  $}

Assume now the case where $F\left(  R,\square R\right)  $ is a separable
function, i.e. $F_{,\chi\zeta}=0$. Hence, by replacing $F\left(  R,\square
R\right)  =f\left(  R\right)  +K\left(  \square R\right)  $ in~$\,$%
(\ref{gup.10}) we derive%
\begin{equation}
S=\int dx^{4}\sqrt{-g}\left[  M\left(  \chi\right)  R-g^{\mu\nu}\chi_{;\mu
}\psi_{;\nu}+V\left(  \chi\right)  +\hat{V}\left(  \psi\right)  \right]
\label{gup11}%
\end{equation}
where $\psi=K_{,\zeta}$, $M\left(  \chi\right)  =f_{,\chi}\left(  \chi\right)
,$ $V\left(  \chi\right)  =f\left(  \chi\right)  -\chi f_{,\chi}\left(
\chi\right)  $ and $\hat{V}\left(  \psi\right)  =K\left(  \zeta\right)  -\zeta
K_{,\zeta}\left(  \zeta\right)  $.

We observe that the Action Integral (\ref{gup11}) is of the form of the GUP
scalar-tensor theory (\ref{bd.08}) for zero value of parameter $\omega,~$that
is, $\omega=0$. Moreover, for $K\left(  \square R\right)  =\frac{1}%
{2\hat{\beta}}\left(  \square R\right)  ^{2}$, we find $\hat{V}\left(
\psi\right)  =-\frac{\hat{\beta}}{2}\psi^{2}$, which means that that the
separable $F\left(  R,\square R\right)  =f\left(  R\right)  +\frac{1}%
{2\hat{\beta}}\left(  \square R\right)  ^{2}$ introduce new geometrodynamical
terms in the field equations which can be attributed to the corrections
provided by the quadratic GUP in scalar-tensor theory. In our case, the
quantum corrections of $f\left(  R\right)  $-gravity are on the limit of
O'Hanlon theory where $\omega=0$. Parameter $\hat{\beta}$ is the analogue of
the deformed parameter $\beta$ of GUP.

In the special case where $f\left(  R\right)  $ is a linear function and for
the $F\left(  R,\square R\right)  =R-2\Lambda+\frac{1}{2\alpha}\left(  \square
R\right)  ^{2}$ theory, from (\ref{gup11}) it follows%
\begin{equation}
S=\int dx^{4}\sqrt{-g}\left[  R-2\Lambda-\hat{\beta}g^{\mu\nu}\chi_{;\mu}%
\psi_{;\nu}-\frac{\hat{\beta}}{2}\psi^{2}\right]  .
\end{equation}

\section{Cosmological solutions}

\label{sec5}

We proceed with the analysis of cosmological evolution for the physical
parameters for the gravitational Action Integral (\ref{gup11}) with background
geometry the FLRW spacetime (\ref{flrw}).

For the spatially flat\ FLRW geometry, the field equations can be derived from
the variation of the point-like Lagrangian function
\begin{equation}
L\left(  a,\dot{a},\chi,\dot{\chi},\psi,\dot{\psi}\right)  =6aM\left(
\chi\right)  \dot{a}^{2}+6a^{2}M_{,\phi}\left(  \chi\right)  \dot{a}\dot{\chi
}+a^{3}\dot{\chi}\dot{\psi}+a^{3}\left(  V\left(  \chi\right)  +\hat{V}\left(
\psi\right)  \right)  , \label{gg1}%
\end{equation}
in which the first modified Friedmann equation is the constraint equation%
\begin{equation}
6M\left(  \chi\right)  H^{2}+6M_{,\chi}\left(  \chi\right)  H\dot{\chi}%
+\dot{\chi}\dot{\psi}-\left(  V\left(  \chi\right)  +\hat{V}\left(
\psi\right)  \right)  =0, \label{gg2}%
\end{equation}
which can be seen as the conservation laws of energy for the three-dimensional
dynamical system described by the point-like Lagrangian (\ref{gg1}). The rest
of the field equations are the Euler-Lagrange equations $\left(  \frac{d}%
{dt}\frac{\partial}{\partial\mathbf{\dot{y}}}-\frac{\partial}{\partial
\mathbf{y}}\right)  L\left(  \mathbf{y,\dot{y}}\right)  =0$, where now
$\mathbf{y=}\left(  a,\chi,\psi\right)  $.

Consequently, the second-order differential equations are%
\begin{equation}
2M\left(  2\dot{H}+3H^{2}\right)  +2M_{,\chi}\left(  2H\dot{\chi}+\ddot{\chi
}\right)  +2\dot{\chi}^{2}M_{,\chi\chi}-\dot{\chi}\dot{\psi}-\left(  V\left(
\chi\right)  +\hat{V}\left(  \psi\right)  \right)  =0, \label{fe.01}%
\end{equation}%
\begin{equation}
\ddot{\chi}+3H\dot{\chi}-\hat{V}\left(  \psi\right)  _{,\psi}=0,
\end{equation}%
\begin{equation}
\ddot{\psi}+3H\dot{\psi}+6M_{,\chi}\left(  \dot{H}+2H^{2}\right)  -V_{,\chi
}=0. \label{fe.03}%
\end{equation}

Let us focus in the case where $M\left(  \chi\right)  =1$ and $V\left(
\chi\right)  =0$. Then the latter gravitational field equations read
\begin{equation}
3H^{2}+\frac{1}{2}\left(  \dot{\chi}\dot{\psi}-\hat{V}\left(  \psi\right)
\right)  =0, \label{fe.04}%
\end{equation}%
\begin{equation}
2\dot{H}+3H^{2}-\frac{1}{2}\left(  \dot{\chi}\dot{\psi}+\hat{V}\left(
\psi\right)  \right)  =0, \label{fe.05}%
\end{equation}%
\begin{equation}
\ddot{\chi}+3H\dot{\chi}-\hat{V}\left(  \psi\right)  _{,\psi}=0, \label{fe.06}%
\end{equation}%
\begin{equation}
\ddot{\psi}+3H\dot{\psi}+6\left(  \dot{H}+2H^{2}\right)  =0. \label{fe.07}%
\end{equation}

Therefore, the effective energy density and pressure components are defined as%
\begin{equation}
\rho_{eff}=\frac{1}{2}\left(  -\dot{\chi}\dot{\psi}+\hat{V}\left(
\psi\right)  \right)  ,
\end{equation}%
\begin{equation}
p_{eff}=-\frac{1}{2}\left(  \dot{\chi}\dot{\psi}+\hat{V}\left(  \psi\right)
\right)  ,
\end{equation}
where the effective equation of state parameter is written
\begin{equation}
w_{eff}=\frac{\left(  \dot{\chi}\dot{\psi}+\hat{V}\left(  \psi\right)
\right)  }{\left(  \dot{\chi}\dot{\psi}-\hat{V}\left(  \psi\right)  \right)
}.
\end{equation}
Indeed, when the potential function $\hat{V}\left(  \psi\right)  $ dominates
the asymptotic solution is that of the de Sitter universe with $w_{eff}=-1$.

If we define the new scalar fields $\chi=\phi+\varpi,~\dot{\psi}=\phi-\varpi$,
then equation (\ref{fe.04}) reads%
\begin{equation}
3H^{2}+\frac{1}{2}\left(  \left(  \dot{\phi}^{2}-\dot{\varpi}^{2}\right)
-\hat{V}\left(  \phi-\varpi\right)  \right)  =0,
\end{equation}
which means that the model is equivalent to the quintom cosmological model
with a mixed potential term, which is different from the analysis presented in
\cite{quin11}. For a complete analysis of the dynamics in quintom cosmology we
refer the reader in \cite{quin22}.

Below we present the phase-space analysis for the field equations
(\ref{fe.04})-(\ref{fe.07}) by using dimensionless variables in the
$H$-normalization approach \cite{q5}.

\subsection{Phase-space analysis}

We define the new dimensionless variables%
\begin{equation}
x=\frac{\dot{\chi}}{H}~,~z=\frac{\dot{\psi}}{6H}\text{ },~w=\frac{\hat
{V}\left(  \psi\right)  }{6H^{2}}~,~\lambda=\frac{\hat{V}_{,\psi}}{\hat{V}%
},~\tau=\ln a
\end{equation}
and we rewrite the field equations (\ref{fe.04})-(\ref{fe.07}) in the form of
the subsequent algebraic-differential system%
\begin{align}
\frac{dx}{d\tau}  &  =\frac{3}{2}\left(  w\left(  4\lambda-x\right)  -x\left(
1+x+z\right)  \right)  ,\\
\frac{dz}{d\tau}  &  =-\frac{3}{2}z\left(  1+w+xz\right)  ,\\
\frac{dw}{d\tau}  &  =3w\left(  1-w-\left(  x-2\lambda\right)  z\right)  ,\\
\frac{d\lambda}{d\tau}  &  =6\lambda^{2}z\left(  \Gamma\left(  \lambda\right)
-1\right)  ~,~\Gamma\left(  \lambda\right)  =\frac{V_{\psi\psi}V}{\left(
V_{,\psi}\right)  ^{2}}\text{\thinspace},
\end{align}
with constraint%
\begin{equation}
1+xz-w=0.
\end{equation}

By utilizing the constraint equation, we can effectively reduce the dimension
of the aforementioned dynamical system by one.

Hence, we arrive at the three-dimensional system%
\begin{align}
\frac{dx}{d\tau}  &  =3\left(  2\lambda-x\right)  \left(  1+xz\right)
,\label{sta1}\\
\frac{dz}{d\tau}  &  =-3z\left(  1+xz\right)  ,\label{sta2}\\
\frac{d\lambda}{d\tau}  &  =6\lambda^{2}z\left(  \Gamma\left(  \lambda\right)
-1\right)  \text{.} \label{sta3}%
\end{align}

At each stationary point $P=\left(  x\left(  P\right)  ,z\left(  P\right)
,\lambda\left(  P\right)  \right)  ~$of the dynamical system (\ref{sta1}%
)-(\ref{sta3}) describes an asymptotic solution where the effective fluid
source has the equation of state parameter%
\begin{equation}
w_{eff}=-1-2x\left(  P\right)  z\left(  P\right)  \text{.} \label{sta4}%
\end{equation}

\subsection{Exponential potential}

In order to reduce further the dimension of the dynamical system we consider
the simple case where $\hat{V}\left(  \psi\right)  $ is the exponential
potential, that is, $\hat{V}\left(  \psi\right)  =V_{0}e^{\lambda_{0}\psi}$.
For this potential we calculate $\Gamma\left(  \lambda\right)  =1$, and
$\lambda=\lambda_{0}$ is always a constant.

Hence, the stationary points of the dynamical system (\ref{sta1}),
(\ref{sta2}) are
\[
P_{1}=\left(  x_{1},-\frac{1}{x_{1}}\right)  ~,~P_{2}=\left(  2\lambda
,0\right)  \text{.}%
\]

For the family of points $P_{1}$ we derive $w_{eff}\left(  P_{1}\right)  =1$,
this implies that the points represent a family of stiff fluid solutions.
Furthermore, the asymptotic solution at point $P_{2}$ describes the
accelerated de Sitter universe, because $w_{eff}\left(  P_{2}\right)  =-1$.

Let us now proceed with the investigation of the stability properties of the
points. For the linearization system (\ref{sta1}), (\ref{sta2}) around the
stationary points $P_{1}$ we determine the eigenvalues $e_{1}\left(
P_{1}\right)  =6\left(  1-\frac{\lambda}{x_{1}}\right)  $,~$e_{2}\left(
P_{1}\right)  =0$. For $1-\frac{\lambda}{x_{1}}>0$ points $P_{1}$ are sources,
however for $1-\frac{\lambda}{x_{1}}<0$ we will infer about the stability of
the points from the phase-space portraits. For point $P_{2}$ the eigenvalues
of the linearized system are $e_{1}\left(  P_{2}\right)  =-3$ and
$e_{2}\left(  P_{2}\right)  =-3$, which means that the de Sitter universe is a
future attractor for the dynamical system.

\begin{figure}[ptb]
\centering\includegraphics[width=1\textwidth]{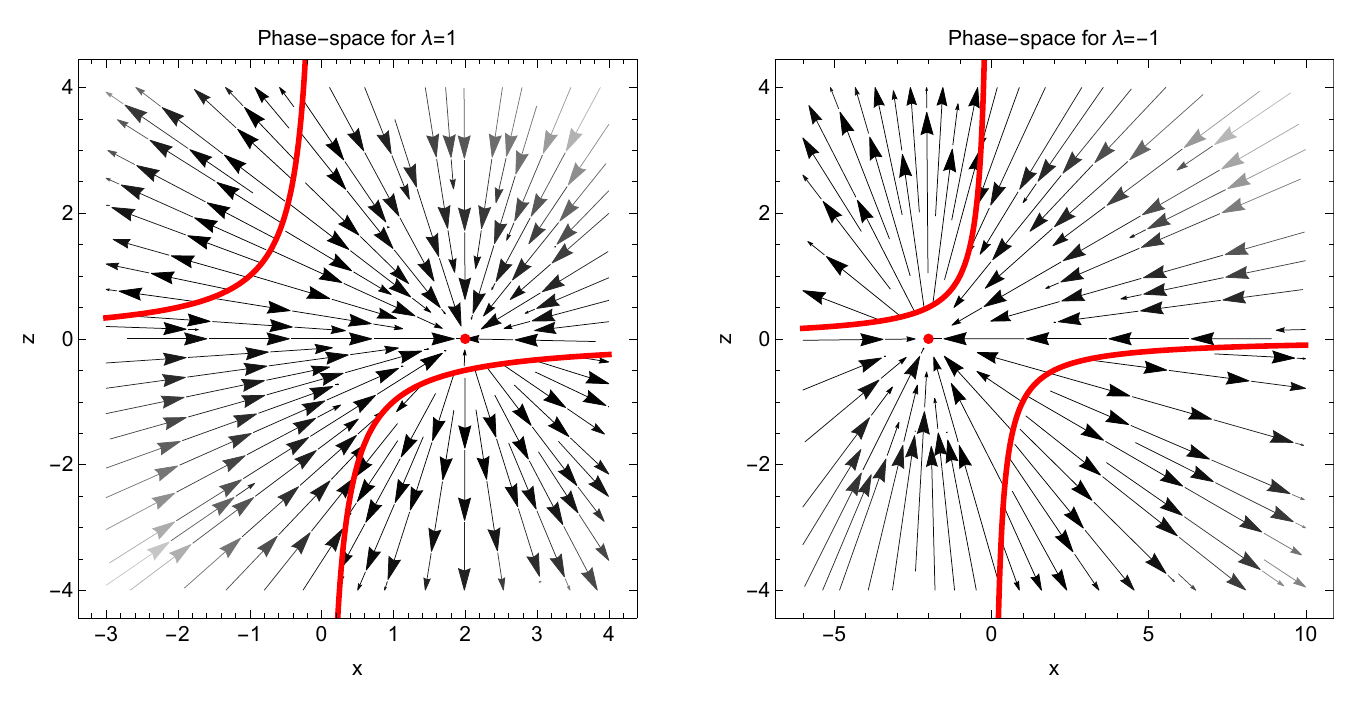}\caption{Phase-space
portrait for the two dynamical system (\ref{sta1}), (\ref{sta2}) for
$\lambda=1$ (left fig.) and $\lambda=-1$ (right fig.). With red lines are the
family of points $P_{1}$ and with red dot is marked the de Sitter attractor
$P_{2}$. We observe that points $P_{1}$ describe always unstable solutions and
are the limits where the trajectories move to the infinity regime. }%
\label{fig1}%
\end{figure}

In Fig. \ref{fig1} we present the phase-space portrait for the two-dimensional
dynamical system (\ref{sta1}), (\ref{sta2}). We observe that the family of
points $P_{1}$ describe always unstable solutions and they are the boundaries
where the trajectories move to the infinity.

\subsubsection{Analysis at infinity}

We define the Poincare variables%
\[
x=\frac{X}{\sqrt{1-X^{2}-Z^{2}}}~,~z=\frac{Z}{\sqrt{1-X^{2}-Z^{2}}}%
\]
and the new independent variable $dT=\sqrt{1-X^{2}-Z^{2}}d\tau$.

In the Poincare variables the field equations read%
\begin{align}
\frac{dX}{dT}  &  =-3\left(  X^{2}+Z^{2}-1-XZ\right)  \left(  2\lambda
-X\left(  2\lambda X+\sqrt{1-X^{2}-Z^{2}}\right)  \right)  ,\\
\frac{dZ}{dT}  &  =3Z\left(  X^{2}+Z^{2}-1-XZ\right)  \left(  2\lambda
X+\sqrt{1-X^{2}-Z^{2}}\right)  ,
\end{align}
and the $w_{eff}\left(  x,z\right)  $ becomes%
\begin{equation}
w_{eff}\left(  X,Z\right)  =-1-\frac{2XZ}{1-X^{2}-Z^{2}}\,.
\end{equation}

Hence, in order the solutions at the infinity to be physical accepted it
should be $XZ\geq0$. For $XZ>0$ at infinity the solutions describe Big Rip
singularities with $w_{eff}\rightarrow-\infty$, while in the limit $XZ=0$, the
de Sitter universe is recovered.

\begin{figure}[ptb]
\centering\includegraphics[width=1\textwidth]{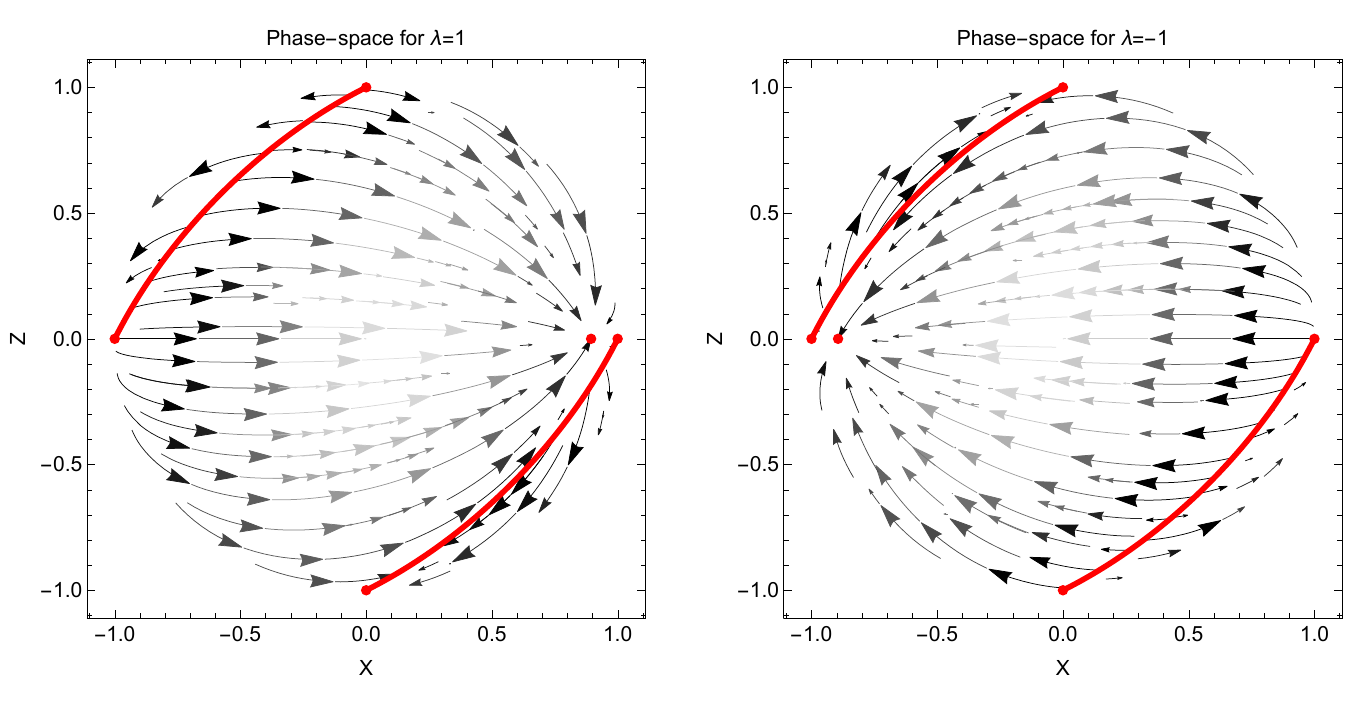}\caption{Phase-space
portrait for the two dynamical system (\ref{sta1}), (\ref{sta2}) for
$\lambda=1$ (left fig.) and $\lambda=-1$ (right fig.) in the Poincare
variables $\left(  X,Z\right)  $. With red lines are the family of points
$P_{1}$ and with red dot are marked the de Sitter attractor $P_{2}$ and the
stationary points at the infinity $Q_{1}^{\pm}$ and $Q_{2}^{\pm}.$ The regions
outside the red lines (points $P_{1}$) lead to unphysical solutions with
$w_{eff}>1,$ thus the accepted initial conditions of the dynamical system are
those inside the red lines. }%
\label{fig2}%
\end{figure}

The stationary points at the infinity are calculated $Q_{1}^{\pm}=\left(
\pm1,0\right)  $ and $Q_{2}^{\pm}=\left(  0,\pm1\right)  $, which means that
$w_{eff}\left(  Q_{1}^{\pm}\right)  =-1$ and $w_{eff}\left(  Q_{1}^{\pm
}\right)  =-1$. The linearized system at points $Q_{1}^{\pm},Q_{2}^{\pm}$ are
$e_{1}\left(  Q_{1}^{\pm}\right)  =0$, $e_{2}\left(  Q_{1}^{\pm}\right)  =0$
and $e_{1}\left(  Q_{2}^{\pm}\right)  =\pm6\lambda_{2}$, $e_{2}\left(
Q_{2}^{\pm}\right)  =0$ respectively. From the phase-space portrait of Fig.
\ref{fig2}, we remark that the stationary points at infinity describe always
unstable solutions and the unique attractor of the dynamical system is the de
Sitter solution described by the stationary point $P_{2}$. The regions outside
the red lines (points $P_{1}$) lead to unphysical solutions with
$w_{eff}\left(  X,Z\right)  >1,$ thus the accepted initial conditions of the
dynamical system are those inside the red lines. 

\section{Conclusions}

\label{sec6}

Scalar field gravitational theories which are modified by the presence of a
minimum length in the Uncertainty Principle are equivalent with two-scalar
field theories of gravity. In this work we investigate the equivalent modified
theory of gravity which can be described by the latter GUP scalar field
theory. We assumed $F\left(  R,\square R\right)  $-gravitational model which
is a sixth-order theory. We found that the separable model $F\left(  R,\square
R\right)  =f\left(  R\right)  +K\left(  \square R\right)  $ can be in
comparison with the GUP scalar field theory. Specifically, for the $K\left(
\square R\right)  =\frac{1}{2\hat{\beta}}\left(  \square R\right)  ^{2}$
model, there are introduced geometrodynamical terms in the field equations
which are in the same form with that introduced by the deformed Heisenberg
algebra of the quadratic GUP. Indeed, parameter $\hat{\beta}$ is linearly
related with the deformation parameter $\beta$ of GUP.

We focused in the case of $F\left(  R,\square R\right)  =R+K\left(  \square
R\right)  $ where now the $K\left(  \square R\right)  $-terms can attribute
the quantum corrections in the limit of General Relativity. For the latter
theory and for a spatially flat FLRW background geometry we investigated the
cosmological dynamics. From the analysis of the phase-space it was found that
there exist always a future attractor which corresponds to the de Sitter
solution without to introduce a cosmological constant term. We conclude that
the quantum corrections related by the $K\left(  \square R\right)  $ can drive
the dynamics such that inflation to occurred.

In a future work we plan to investigate further the effects of the
introduction of higher-order terms $K\left(  \square R\right)  $ in $f\left(
R\right)  $f theory for the cosmological evolution and investigate the case of
EUP in gravitational models.

\textbf{Data Availability Statements:} Data sharing is not applicable to this
article as no datasets were generated or analyzed during the current study.

\begin{acknowledgments}
The author thanks the support of Vicerrector\'{\i}a de Investigaci\'{o}n y
Desarrollo Tecnol\'{o}gico (Vridt) at Universidad Cat\'{o}lica del Norte
through N\'{u}cleo de Investigaci\'{o}n Geometr\'{\i}a Diferencial y
Aplicaciones, Resoluci\'{o}n Vridt No - 098/2022.
\end{acknowledgments}


\begin{thebibliography}{999}                                                                                              %


\bibitem {rr1}A.G. Riess et al., Astron. J. 116, 1009 (1998)

\bibitem {Teg}M. Tegmark et al., Astrophys. J. 606, 702 (2004)

\bibitem {Kowal}M. Kowalski et al., Astrophys. J. 686, 749 (2008)

\bibitem {Komatsu}E. Komatsu et al., Astrophys. J. Suppl. Ser. 180, 330 (2009)

\bibitem {suzuki11}N. Suzuki et al., Astrophys. J. 746, 85 (2012)

\bibitem {l01}S. Weinberg, Rev. Mod. Phys. 61, 1\ (1989)

\bibitem {l02}T. Padmanabhan, Phys. Rept. 380, 235 (2003)

\bibitem {l03}L. Perivolaropoulos and F. Skara, New Astronomy Reviews, 95,
101659 (2022)

\bibitem {ratra}P. Ratra and L. Peebles, Phys. Rev. D 37, 3406 (1988)

\bibitem {q5}E.J. Copeland, M. Sami and S.\ Tsujikawa, Int. J. Mod. Phys. D
15, 1753 (2006)

\bibitem {qq1}C. Rubano and J.D. Barrow, Phys. Rev. D 64, 127301 (2001)

\bibitem {qq4}S. Basilakos and G. Lukes-Gerakopoulos, Phys.\ Rev. D 78, 083509 (2008)

\bibitem {q22}R.C. Caldwell, M. Kamionkowski and N.N. Weinberg, Phys. Rev.
Lett. 91, 071301 (2003)

\bibitem {q23}F. Briscese, E. Elizalde,\ S. Nojiri and S.D. Odintsov, Phys.
Lett. B 646, 105 (2007)

\bibitem {qq01}J. Khoury and A. Wetlman, Phys.\ Rev. D 69, 044026 (2004)

\bibitem {st1}J.E.M. Aguilar, M. Montes and A. Bernal, Phys. Scr. 98, 035021 (2023)

\bibitem {st2}C.H.-T. Wang, J.A. Reid, A.St.J. Murphy, D. Rodrigues, M. Al
Alawi, R. Bingham, J.T. Mendonca and T.B. Davies, Phys. Lett. A\ 380, 3761 (2016)

\bibitem {st3}A. Zucca, L. Pogosian, A. Silvestri, Y. Wang and G.-B. Zhao,
Phys. Rev. D 101, 043518 (2020)

\bibitem {st4}G. Koutsoumbas, K. Ntrekis, E. Papantonopoulos and E.N.
Saridakis, JCAP 18, 003 (2018)

\bibitem {st5}R.\ Gannouji and M. Sami, Phys. Rev. D 82, 024011 (2010)

\bibitem {st6}A. Paliathanasis, Class. Quantum Grav. 37, 195014 (2020)

\bibitem {st7}G. Leon, A. Paliathanasis and J.L. Morales-Martinez, Eur. Phys.
J. C 78, 753 (2018)

\bibitem {st8}T. Pati, S. Panda, M.\ Sharma and Ruchika, Eur. Phys. J. C 83,
131 (2023)

\bibitem {q20}A. Al Mamon, A. Paliathanasis and S. Saha, l, Eur. Phys. J. C
82, 232 (2021)

\bibitem {q21}S. Basilakos, G. Lukes-Gerakopoulos, Phys.\ Rev. D 78, 083509 (2008)

\bibitem {q21a}H. Li, W. Yang and L.\ Gai, A\&A 623, A28 (2019)

\bibitem {q21b}V. A. Popov, Phys. Lett. B 686, 211 (2010)

\bibitem {q21c}R. von Marttens, D. Babosa and J. Alcaniz, JCAP 04, 052 (2023)

\bibitem {q26}C. Deffayet, X. Gao, D.A. Steer and G.\ Zahariade, Phys.\ Rev. D
84, 064039\ (2011)

\bibitem {q26a}B. Chen, Y. Wu, J. Chi, W. Liu and Y. Hu, Universe 8, 520 (2022)

\bibitem {q25}A. Kundu, S.D. Pathak and V.K. Ojha, Comm. Theor. Phys. 73,
025402 (2021)

\bibitem {q25a}J.S.Bagla, H.K.Jassal and T.Padmanabhan, Phys. Rev. D 67,
063504 (2003)

\bibitem {q25b}E. Nouri, H. Motavalli and A.R. Akbarieh, Int. J. Mod. Phys. D
31, 2250020 (2022)

\bibitem {md1}S. Nojiri, S.D. Odintsov and V.K. Oikonomou, Phys.Rept. 692, 1 (2017)

\bibitem {md2}A. Joyce, L. Lombriser and F. Schmidt, Annu. Rev. Nucl. Part.
Sci. 66, 122 (2016)

\bibitem {md3}T. Clifton, P.G. Ferreira, A. Padilla and C. Skordis, Phys.
Rept. 513, 1 (2012)

\bibitem {cp1}S. Capozziello and V. Faraoni, Gen. Relativ. Grav. 40, 357 (2008)

\bibitem {Buda}H.A. Buchdahl, Mon. Not. Roy. Astron. Soc. 150, 1 (1970)

\bibitem {Ferraro}R. Ferraro and F. Fiorini, Phys. Rev. D 75, 084031 (2007)

\bibitem {f6}J. B. Jimenez, L. Heisenberg and T. Koivisto, Phys. Rev. D 98,
044048 (2018)

\bibitem {bb1}G. Cognola, E. Elizalde, S. Nojiri, S.D. Odintsov and S.
Zerbini, Phys. Rev. D 73, 084007 (2006)

\bibitem {bb2}G. Kofinas and E.N.\ Saridakis, Phys. Rev. D 90, 084044 (2014)

\bibitem {bb3}T. Harko, F.S.N. Lobo, G. Otalora and E.N. Saridakis, JCAP 12,
021 (2014)

\bibitem {bb4}A. Paliathanasis and G. Leon, Eur. Phys. J. Plus 136, 1092 (2021)

\bibitem {bb5}H.R. Kausar, R. Saleem and A. Ilyas, Phys. Dark Universe 26,
100401 (2019)

\bibitem {bb6}G.J. Olmo, Int. J. Mod. Phys. D 20, 413 (2011)

\bibitem {bb7}A. De Felice, J.-M. Gerard and T. Suyama, Phys. Rev. D 82,
063526 (2010)

\bibitem {Sotiriou}T.P. Sotiriou and V. Faraoni Rev. Mod. Phys. 82 451 (2010)

\bibitem {odin1}S. Nojiri and S.D. Odintsov, Phys. Rep. 505 59 (2011)

\bibitem {cf1}V. K. Oikonomou and I. Giannakoudi, Int. J. Mod. Phys. D 31,
2250075 (2022)

\bibitem {cf2}S.P. Hatkar, P.S. Dudhe and S.D. Katore, Foundations in Physics
49, 1067 (2019)

\bibitem {cf3}S. Nojiri and S. Odintsov, Int. J. Geom. Meth. Mod. Phys. 11,
1460006 (2014)

\bibitem {cf4}A. de la Cruz-Dombriz, A. Dobado and A. L. Maroto, Phys. Rev. D
80, 124011 (2009)

\bibitem {cf5}Kh Jafarzade, E Rezaei and S H Hendi, Progress of Theoretical
and Experimental Physics 2023, 053E01 (2023)

\bibitem {cf6}G.\thinspace G.\thinspace L. Nashed and S. Nojiri, Phys. Rev. D
102, 124022 (2020)

\bibitem {qua1}H. Nariai and K. Tomita, Prog. Theor. Phys. 46, 776 (1971)

\bibitem {qua2}G.V. Bicknell, J. Phys. A.: Math. Nucl. Gen. 7, 1061 (1974)

\bibitem {qua3}J.D. Barrow, Nucl. Phys. B 296, 679 (1988)

\bibitem {planck2015}P.A.R. Ade et al. (Planck 2015 Collaboration), A.\&A.
594, A20 (2016)

\bibitem {star}A.A. Starobinsky, Phys. Lett. B 91, 99 (1980)

\bibitem {df1}H. Nariai, Prog. Theor. Phys. 46, 433 (1971)

\bibitem {df2}F.C. Michel, Ann. Phys. 76, 281 (1973)

\bibitem {Maggiore}M. Maggiore, Phys. Lett. B 304, 65 (1993).

\bibitem {cas1}R. Casadio and F. Scardigli, Phys. Lett. B 807, 135558 (2020)

\bibitem {cas2}S. Benczik, L.N. Chang, D. Minic, N. Okamura, S. Rayyan and T.
Takeuchi, Phys. Rev.D 66, 026003 (2002)

\bibitem {cas3}A.F. Ali, Class. Quantum Grav. 28, 065013 (2011)

\bibitem {mc1}E.C. Vagenas, A.F. Ali, M. Hemeda and H. Alshal, Eur. Phys. J. C
79, 398 (2019)

\bibitem {mc2}A.F. Ali, S. Das and E.C. Vagenas, Phys. Lett. B 678, 497 (2009)

\bibitem {mc3}Z. Silagadge, Phys. Lett. A 373, 2643 (2009)

\bibitem {cc1}Y.-G. Miao and Y.-J. Zhao, Int. J. Phys. D. 23, 1450062 (2014)

\bibitem {cc2}L.N. Chang, D. Minic, N. Okamura and T. Takeuchi, Phys. Rev D
65, 125028 (2002)

\bibitem {ep1}C. Bambi and F.R. Urban, Class. Quantum Grav. 25, 095006 (2008)

\bibitem {ep2}S. Aghababaei, H. Moradpour and E.C. Vagenas, Eur. Phys. J. Plus
136, 997 (2021)

\bibitem {gup11}P. Bosso, G.G. Luciano, L. Petruzziello and F. Wagner, 30
years in: Quo vadis generalized uncertainty principle? (2023) [arXiv:2305.16193]

\bibitem {angup1}A.~Paliathanasis, S.~Pan and S.~Pramanik, Class. Quant. Grav.
32, no.24, 245006 (2015)

\bibitem {angup2}A. Giacomini, G. Leon, A. Paliathanasis and S. Pan, Eur.
Phys. J. C 80, 931 (2020)

\bibitem {angup3}A. Paliathanasis, G. Leon, W. Khyllep, J. Dutta and S. Pan,
Eur. Phys. J. C 81, 607 (2021)

\bibitem {angup4}A. Paliathanasis and G. Leon, Gen. Relativ.\ Grav. 55, 12 (2023)

\bibitem {ohan}J. O'Hanlon, Phys. Rev. Lett. 29, 137 (1972)

\bibitem {far1}V. Faraoni, Cosmology in Scalar-Tensor Gravity. Dordrecht,
Kluwer Academic (2004)

\bibitem {brans}C.H Brans and R.H. Dicke, Phys. Rev. 124, 925 (1961)

\bibitem {lan1}C. Gao, Y. Gan, X. Wang and X. Chen, Phys. Lett. B 702, 107 (2011)

\bibitem {lan2}S. Nojiri, S.D. Odintsov, V.K. Oikonomou and T. Paul, Phys.
Rev. D 100, 084056 (2019)

\bibitem {lan3}D. F. Jimenez, L. N. Granda, and E. Elizalde, Int. J. Mod.
Phys. D 28, 1950171 (2019)

\bibitem {sx1}S. Gottlober, H.-J. Schmidt and A.A. Starobinsky, Class. Quamtum
Grav. 7, 893 (1990)

\bibitem {sx2}D. Wands, Class. Quantum Grav. 11, 269 (1994)

\bibitem {sx3}A.L. Berkin and K.-i. Meida, Phys. Lett. B 245, 348 (1990)

\bibitem {sx4}A.R.R. Castellanos, F. Sobreira, I.L. Shapiro and A.A.
Starobinsky, JCAP 12, 007 (2018)

\bibitem {sx4a}L. Amendola, A.B. Mayer, S.\ Capozziello, F. Occhionero and S.
Gottlober, Class. Quantum Grav. 10, L43 (1993)

\bibitem {sx5}S. Carloni, J.L. Rosa and J.P.S. Lemos, Phys. Rev. D 99, 104001 (2019)

\bibitem {sx6}I. Quandt and H.-J.\ Schmidt, Astron. Nachr. 312, 97 (1991)

\bibitem {sx7}S. Nesseris and A. Mazumbar, Phys. Rev.\ D 79, 104006 (2009)

\bibitem {sx8}M. Skugoreva, A.\ Toporensky and P.\ Tretyakov, Grav. Cosmol.
17, 110 (2011)

\bibitem {sx9}M. Iihoski, JCAP 02, 022 (2011)

\bibitem {sx10}S. Rani, M.B.A.\ Sulehri and A. Jawad, Phys. Dark Univ. 29,
100555 (2020)

\bibitem {Vagenas}S. Das and E.C. Vagenas, Phys. Rev. Lett. 101, 221301 (2008)

\bibitem {neg1}P. Jizba, H. Kleinert and F. Scardigli, Phys. Rev. D 81, 084030 (2010)

\bibitem {neg2}L. Buoninfante, G.G. Luciano and L. Petruzziello, Eur. Phys. J.
C 79, 663 (2019).

\bibitem {neg5}L. Buoninfante, G. Lambiase, G.G. Luciano and L. Petruzziello,
Eur. Phys. J. C 80, 853 (2020)

\bibitem {neg6}S. Giardino and V. Salzano, Eur. Phys. J. C 81, 110 (2021)

\bibitem {Kemph1}A. Kempf, J. Phys. A: Math. Gen. 30, 2093 (1997)

\bibitem {Kemph2}H. Hinrichen and A. Kempf, J. Math. Phys. 37, 2121 (1996)

\bibitem {Moayedi}S.K. Moayedi, M.R. Setare and H Moayeri, Int. J. Theor.
Phys. 49, 2080 (2010)

\bibitem {Tikhonov}A.N. Tikhonov, Sbornic. Math. 31, 575 (1952)

\bibitem {quin11}E.N. Saridakis, J.M. Weller, Phys. Rev. 81, 123523 (2010)

\bibitem {quin22}G. Leon, A. Paliathanasis and J.L. Morales-Martinez, Eur.
Phys. J. C 75, 753 (2018)
\end{thebibliography}
\end{document}